\documentclass[10pt]{article}
\usepackage[]{times,emulateapj}

\def\hh {\mathrm{H}_2}               
\def\ee #1 {\times 10^{#1}}          
\def\ut #1 #2 { \, \mathrm{#1}^{#2}} 
\def\u #1 { \, \mathrm{#1}}          
\newcommand{\msun}{\mathrm{M_{\sun}}}

\def\tkh {t_{\mathrm{KH}}}

\slugcomment{Ap. J. Lett., in press}

\lefthead{Wardle \& Walker}
\righthead{Cold Clouds in Galaxy Halos}

\begin{document}

\title{Thermal Stability of Cold Clouds in Galaxy Halos}

\author{Mark Wardle \& Mark Walker}
\affil{Special Research Centre for Theoretical Astrophysics, \\
School of Physics, University of Sydney, NSW 2006, Australia}


\begin{abstract} 
We consider the thermal properties of cold, dense clouds of molecular 
hydrogen and atomic helium.  For cloud masses below 
$10^{-1.7} \msun$, the internal pressure is sufficient to permit 
the existence of particles of solid or liquid hydrogen at temperatures 
above the microwave background temperature.  Optically thin thermal 
continuum emission by these particles can balance cosmic-ray heating 
of the cloud, leading to equilibria which are thermally {\it stable\/} 
even though the heating rate is independent of cloud temperature.  For 
the Galaxy, the known heating rate in the disk sets a minimum mass of 
order $10^{-6}\msun$ necessary for survival.  Clouds of this 
type may in principle comprise most of the dark matter in the Galactic 
halo.  However, we caution that the equilibria do not exist at 
redshifts $z\ga1$ when the temperature of the microwave background was 
substantially larger than its current value; the formation and 
survival of such clouds to the present epoch therefore remain open 
questions.
\end{abstract}



\section{Introduction}

Walker \& Wardle (1998) showed that a population of neutral, AU-sized
clouds in the Galactic halo could be responsible for the ``Extreme
Scattering Events'' (ESEs) observed in the radio flux towards several
quasars (Fiedler et al.  1987, 1994).  In this model the cloud
surfaces are exposed to UV radiation from hot stars in the Galactic
disk, producing a photo-ionised wind.  When one of these clouds
crosses the line of sight to a compact radio source, the flux varies
as a result of refraction by the ionised gas (cf.  Henriksen \& Widrow
1995).  This model explains the observed flux variations quite
naturally; but if the clouds are self-gravitating, then the ESE event
rate implies that the cloud population comprises a significant
fraction of the Galaxy's mass.

This halo cloud population cannot contain much dust mixed with the gas
as this would lead to optical extinction events of distant
stars: either the clouds have extremely low metallicity, or any dust
grains have sedimented to the cloud centre.  Given this, several
factors make the clouds difficult to detect (Pfenniger, Combes
\& Martinet 1994): cold molecular hydrogen is, by and large,
invisible; the clouds are small; they are transparent in most regions
of the electromagnetic spectrum; and they cover a small fraction of the
sky. The clouds are not sufficiently compact to cause gravitational
lensing towards the LMC, although Draine (1998) has shown that there
is substantial optical refraction by the neutral gas, so that
microlensing experiments (Paczy\'nski 1996) already place useful
constraints on the properties of low-mass halo clouds.

Given that this hypothesised cloud population does not violate
observational constraints, the primary issues that need to be
addressed are theoretical: (i) how and when did these clouds form?
and (ii) how do they resist gravitational collapse?  The second of
these is addressed in this \emph{Letter}.

We begin by writing down equations describing a simple ``one-zone''
model of a cloud, characterised by a single temperature and
pressure (\S\ref{sec:model}), and show that particles of solid H$_2$ may
exist in the clouds (cf. Pfenniger \& Combes 1994).  At temperatures
above the microwave background temperature, these particles cool the
cloud by thermal continuum radiation, admitting equilibria in which this
cooling balances heating by cosmic rays.  In \S\ref{sec:stability}, we
show that (for optically thin emission) these equilibria are thermally
stable: if the cloud contracts the coolant is destroyed by the increase
in temperature, and the power deposited by cosmic rays causes the cloud
to expand and the temperature to return to its original value. We conclude
that, within the context of our one-zone model, the viable mass range for
Galactic clouds is $10^{-6}$--$10^{-1.7}\msun$.

\section{Cloud model}
\label{sec:model}

Virial equilibrium implies that for a self-gravitating cloud
characterised by mass $M$, temperature $T$, and radius $R$,
\begin{equation}
	R \approx GM\mu/kT
	\label{eq:R}
\end{equation}
where $\mu$ is the mean molecular weight.
The pressure in the cloud can be related to the temperature upon
noting that $P\sim GM^2/R^4$, yielding
\begin{equation}
	P = \frac{q}{G^3M^2} \, \left( \frac{kT}{\mu} \right)^4 \,,
	\label{eq:P}
\end{equation}
where $q$ depends on the cloud's structure.  For polytropes $q$ rises
monotonically from 9.2 to 40 as the polytropic index runs from 3/2
to 9/2, so we adopt $q=20$.


\begin{figure*}   
\centerline{\epsfxsize=8cm \epsfbox{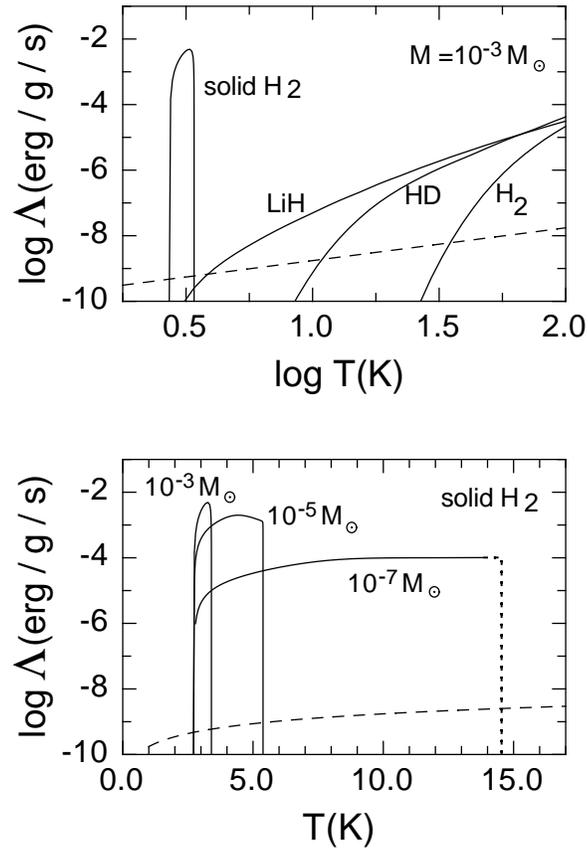}}
\figcaption{\emph{Top:} cooling by rotational transitions of H$_2$, HD 
and LiH in the gas phase, and thermal continuum emission from 
particles of solid H$_2$ for a $10^{-3}\msun$ cloud in virial 
equilibrium.  \emph{Bottom:} solid H$_2$ cooling rates for cloud 
masses $10^{-3}$, $10^{-5}$ and $10^{-7}\msun$.  Dotted portion of the 
$10^{-7}\msun$ curve indicates where the coolant is liquid H$_2$.  The 
dashed line in both plots indicates the cooling rate that would give a 
Kelvin-Helmholtz time scale of 10 Gyr.
\label{fig:cooling}}
\end{figure*}

At sufficiently high pressures a fraction $x$ of the molecular
hydrogen assumes solid (or liquid) form. 
Then $\mu=(1-x+2y)m/(1-x+y)$, where $m$
is the mass of an H$_2$ molecule, and $y\approx1/6$ is the
abundance ratio He:H$_2$ by number.
Neglecting the temperature difference between the
phases, in equilibrium the partial pressure of H$_2$ equals the
saturated vapour pressure, i.e.
	\begin{equation}
		\frac{1-x}{1-x+y}P = \left(\frac{2\pi m}{h^2}\right)^{3/2}
		(kT)^{5/2} e^{-T_v/T} \,,
		\label{eq:Psat}
	\end{equation}
(valid for $0<x<1$), where $k T_v$ is the heat of vapourisation for
H$_2$ (Phinney 1985).  With $T_v = 91.5 \u K $, the vapour
pressure given by the RHS of eq. (\ref{eq:Psat}) is within 20 \% of
the available experimental data (Souers 1986).

Hydrogen grains can cool the gas in a manner similar to dust
grains in molecular clouds: the gas cools via collisions with slightly
colder solid particles, which in turn cool by thermal continuum
emission.  To calculate the cooling by solid H$_2$, first consider the
net power radiated by a single particle (we employ an `escape probability'
formulation of radiative transfer):
\begin{equation}
	L_s = 4\sigma\left(\frac{C(T)T^4}{1+\tau} -
   \frac{C(T_b)T_b^4}{1+\tau_b}\right) \,,
	\label{eq:L_s}
\end{equation}
where $C(T)$ is the Planck-mean absorption cross-section, $T_b$ is the
cosmic microwave background temperature, and $\tau$ and $\tau_b = \tau
C(T_b)/C(T)$ are the Planck-mean optical depth of the cloud to thermal
radiation characterized by $T$ and $T_b$ respectively.  Assuming that
the particle size is $\ll \lambda $ ($ \sim 0.1 \u cm $ at the
temperatures of relevance here), we may write $C(T) = C_m(T) m_s$,
$m_s$ being the particle mass, and for spherical grains we have
(Draine \& Lee 1984)
\begin{equation}
	C_m(T) \approx \frac{15 (4\pi)^3}{28\rho_s}
	\frac{\lambda_2}{(\varepsilon_1+2)^2}
	\left(\frac{kT}{hc}\right)^2\, ,
	\label{eq:Cm}
\end{equation}
where $\rho_s = 0.087 \u g \ut cm -3 $ is the density of
solid H$_2$  (Souers 1986),the complex dielectric function of the
solid is $\varepsilon_1 + i\varepsilon_2$, and we have assumed that
$\varepsilon_2 =
\lambda_2/\lambda$ as expected at low frequencies. The net cooling rate
per unit mass of cloud material (gas and solid) is then
\begin{equation}
	\Lambda = \frac{4\pi R^2\sigma}{M} \left(\frac{\tau T^4}{1+\tau} -
    \frac{\tau_b T_b^4}{1+\tau_b}\right) \,,
	\label{eq:Lambda}
\end{equation}
where $\tau_b = \tau\,(T_b/T)^2$, and
\begin{equation}
	\tau = C_m \frac{x}{1+2y}\, \frac{M} {\pi R^2} \,.
	\label{eq:tau}
\end{equation}

To evaluate $\Lambda$, we require optical constants for solid $\hh$ in 
the microwave.  The particles are expected to be almost pure 
para-hydrogen as an ortho-para mixture of the solid relaxes to para 
($J=0$) form in a few days (Souers 1986).  The low frequency value of 
$\varepsilon_1$ for para-hydrogen has been measured (Souers 1986) as 
$\varepsilon_1\simeq1.25$; the low frequency limit of $\varepsilon_2$ 
is less certain.  Jochemsen et al.\ (1978) measured the extinction 
coefficient of a single crystal of solid para-hydrogen in the region 
of interest ($\lambda\sim0.1\;{\rm cm}$), but could not determine 
whether this continuum extinction was due to absorption or scattering 
within the crystal.  Because these measurements do not conform to the
anticipated low-frequency behaviour ($\propto1/\lambda$), and absorption
bands are not expected below the S(0) line, it is likely that the
absorption of pure crystalline $\hh$ is much smaller than the measured
extinction, and we can only infer a limit: $\varepsilon_2\la 1.8\times10^{-3}$.
However, the low-frequency absorption of solid H$_2$ grains could be
strongly enhanced by impurity species and lattice defects. For the
purposes of this paper we adopt $\varepsilon_2 = \lambda_2/\lambda$
and $\lambda_2 = 10^{-4} \u cm $.  Within the confines of the model,
this assumption represents one of our main areas of  uncertainty.  


\begin{figure*}  
	\centerline{\epsfxsize=8cm \epsfbox{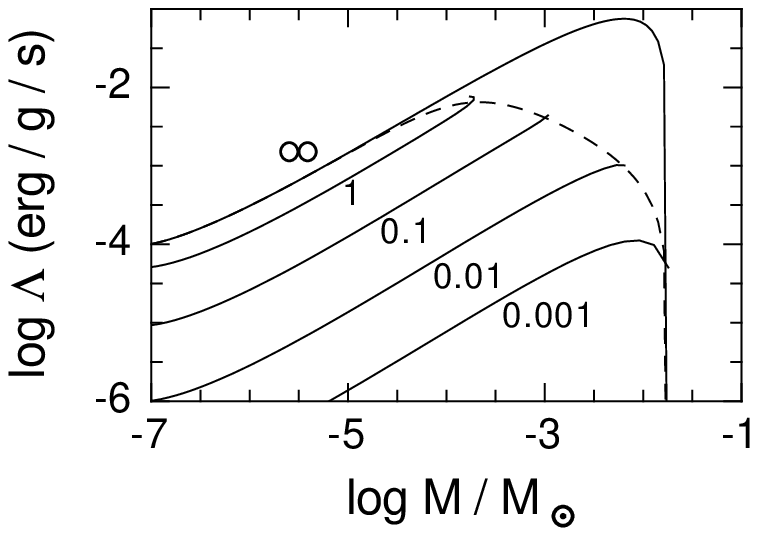}}
	\figcaption{Solid $\hh$ cooling rate versus cloud mass for 
	different optical depths.  The curves, labelled by $\tau$, are 
	truncated at high masses when the solutions become thermally 
	unstable.  The optically thick limit, corresponding to black-body 
	emission, is labelled ``$\infty$''.  The dashed curve 
	delineates where half of the hydrogen is in solid form, an 
	approximate upper limit for thermal stability.
	\label{fig:tau}}
\end{figure*}

For a given cloud mass, the fraction of H$_2$ in the solid phase can 
be determined from $T$ using eqs (\ref{eq:P}) and (\ref{eq:Psat}).  
This allows the cooling by solid particles to be calculated as a 
function of $T$.  The upper panel of Fig \ref{fig:cooling} illustrates 
this for a cloud of mass $10^{-3}\msun$.  For comparison, the cooling 
contributed by rotational lines of gas-phase H$_2$, HD and LiH is also 
plotted; energies and A-values from Turner, Kirby-Docken \& Dalgarno 
(1977), Abgrall, Roueff \& Viala (1982), and Gianturco et al.\ (1996).  
The adopted deuterium and LiH abundances are $3\times10^{-5}$, and 
$1.2 \times 10^{-10}$ respectively (Schramm \& Turner 1998).  We 
employ an escape-probability formulation of radiative transfer, in 
which the optically-thin cooling rates are divided by $(1+\tau)$ where 
$\tau$ is the optical depth at line centre (rotational transitions) or 
the Planck-mean (continuum emission).

There is a critical temperature $T_c$ at which H$_2$ in the cloud lies on 
the border between the solid and gaseous phases, i.e.  the partial 
pressure of H$_2$ is equal to its saturated vapour pressure.  From 
this point $x$ and $\Lambda$ increase precipitously as $T$ is reduced, 
until $\tau \sim 1$ and the cooling is then roughly black-body.  
$\Lambda$ then increases more slowly, peaking when $x\sim 1/2$ and 
subsequently dropping to zero as the temperature approaches that of 
the microwave background.  At lower temperatures the cloud is heated 
by the background radiation.

The solid H$_2$ cooling curves for cloud masses of $10^{-3}$, 
$10^{-5}$ and $10^{-7} \msun$ are compared in the lower panel of 
figure \ref{fig:cooling}.  Decreasing the cloud mass has two 
consequences: the critical temperature $T_c$ decreases, whereas the maximum 
value of $\Lambda$ decreases (because in this circumstance the 
emission is optically thick, and cloud surface area is proportional to 
$M^2$ at a given temperature).  For cloud masses $\la 10^{-7} \msun$, 
$T_c$ is above the H$_2$ triple point (13.8\,K) and liquid droplets of 
H$_2$ form instead.  This does not qualitatively affect the 
calculations as the density and saturated vapour pressure of the 
liquid are within 50\% of those of the solid for $T\la 20 \,K $
(Souers 1986), and the optical properties are similar (in the sense 
that $\varepsilon_2$ is small and uncertain), thus in Fig.  
\ref{fig:cooling} we continue the cooling curve for a $10^{-7}\msun$ 
cloud above the triple point as a dotted curve.  Thermally stable 
solutions do not exist for masses $\la 10^{-7.5}\msun $, as the 
partial pressure of H$_2$ exceeds the saturated vapour pressure unless 
$x > 0.5$ (see \S\ref{sec:stability}).  On the other hand, for $M\ga 
10^{-1.7} \msun$ $T_c$ is below the CMB temperature and the solid phase 
warms the cloud rather than cooling it.

\section{Thermal Equilibrium}
\label{sec:stability}

In thermal equilibrium, the cloud temperature is set by the balance 
between cooling and heating.  We assume that clouds in the Galactic 
halo are heated primarily by cosmic rays.  The local interstellar 
cosmic-ray ionisation rate in the Galactic disc, $\sim 3\ee -17 \ut s 
-1 \ut H -1 $ (Webber 1998), implies a heating rate $\Gamma \sim 
3\times10^{-4} \mathrm{erg\,g^{-1}\,s^{-1}}$ (Cravens \& Dalgarno 
1978).  The cosmic-ray heating in the halo is uncertain but should be 
somewhat lower, say $\sim 10^{-5}\mathrm{erg\,g^{-1}\,s^{-1}}$.  In 
Fig.  \ref{fig:tau}, we show the cooling rates for clouds of masses 
$10^{-7}$--$10^{-1.7}\msun$.  The solid curves are contours of constant 
optical depth; the dashed curve shows the optically-thick limit and 
represents the maximum cooling rate for each cloud mass.  It appears 
that solid hydrogen can provide the necessary cooling for 
planetary-mass gas clouds at the cosmic-ray heating rates expected in 
the Galactic disk and halo.

The upper panel of Fig.  \ref{fig:cooling} shows that there are
typically three equilibrium temperatures available for cloud masses
between $10^{-7.5}$ and $10^{-1.7}\msun$ at the expected heating rates:
solid H$_2$ provides one barely above $T_b$ and one a few degrees
higher; the gas-phase coolants provide an equilibrium above 30 K. We
now show that thermal stability requires $\Lambda$ to be a decreasing
function of $T$, and therefore only the second of these three
equilibria is stable.

In virial equilibrium the total energy per unit mass is approximately 
$-\frac{3}{2}kT/\mu$ (the internal excitation of the gas is 
negligible at the low temperatures of interest here), so the thermal 
evolution of the cloud is determined by
\begin{equation}
	\frac{3k}{2\mu}\frac{{\rm d}T}{{\rm d}t} =\Lambda - \Gamma.
	\label{eq:energy}
\end{equation}
In the absence of heating the cloud contracts on the Kelvin-Helmholtz
time-scale $\tkh = \frac{3}{2} kT / (\mu\Lambda) $.  Note that this
time-scale can be a substantial fraction of the Hubble time for temperatures
of a few Kelvin: the dashed curves in figure \ref{fig:cooling}
show the cooling rate that yields $\tkh = 10 \u Gyr $.  In thermal
equilibrium cosmic-ray heating replaces the energy radiated away by
the cloud implying, for example, $\tkh \sim2\times10^6$~yr for $\Gamma
\sim 10^{-5} \u erg \ut g -1 \ut s -1 $ (at $T\sim10$~K).
This is much greater than
the sound crossing time ($\sim10^2$~yr), so the response of a cloud
to dynamical perturbations is adiabatic, to a good approximation, and
dynamical stability is assured.  However, eq.  (\ref{eq:energy}) shows
that perturbations to the
cloud temperature grow or decay as $e^{\alpha t}$ where
\begin{equation}
	\alpha= \tkh^{-1} \frac{T}{\Lambda}\,\frac{\rm d\;\;}{{\rm d}T}
	(\Lambda - \Gamma),
	\label{eq:growth_rate}
\end{equation}
and the right-hand side of this equation is evaluated at the
equilibrium temperature.
Thus a cloud is thermally stable only if a decrease (increase) in
cloud temperature leads to cooling outstripping (lagging) heating.
For cosmic ray heating, $\Gamma$ is independent of $T$ if the column
through the cloud is insufficient to cause significant attenuation of
cosmic rays (changes in temperature affect the cloud's column density
through the virial relationship $R\propto 1/T$).  Thermal stability
then requires that $\Lambda$ be a decreasing function of $T$, and
we conclude that {\it only the equilibrium on the high-temperature shoulder
of the solid hydrogen cooling curve is stable.\/}
In fact the column density of each cloud ($\sim10^2\;{\rm g\,cm^{-2}}$:
Walker 1999) is sufficient to stop sub-GeV cosmic-ray protons (and
all electrons), leading to a dependence of $\Gamma$ on $T$;
this dependence is too weak to affect our conclusions concerning
stability.

\section{Discussion}
\label{sec:discussion}

The suggestion that cold gas could comprise a significant fraction of 
the Galaxy's dark matter is not new, previous proposals include: a 
fractal medium in the the outer reaches of the Galactic disk 
(Pfenniger et al.\ 1994); isolated halo clouds (Gerhard \& Silk 1996); 
and mini clusters of clouds in the halo (de Paolis et al.\ 1995; Gerhard 
\& Silk 1996).  However, to date there has been no compelling reason 
to believe that isolated, cold gas clouds -- as inferred by Walker \& 
Wardle (1998) -- could support themselves for long periods against 
gravitational collapse.  We have shown that such clouds can be 
stabilised by the precipitation/sublimation of particles of solid 
hydrogen (or by the condensation/evaporation of droplets of liquid 
hydrogen) if these particles dominate the radiative cooling of the 
cloud.  The key feature which confers thermal stability is that these 
particles are destroyed, hence cooling becomes less efficient, as the 
cloud temperature increases.  This feature will be present in any 
model where condensed hydrogen is the principal coolant, and consequently 
we expect that more sophisticated structural treatments will also 
admit stable solutions.

The masses of thermally stable clouds lie in the approximate range 
$10^{-7.5}$--$10^{-1.7} \msun$.  The lower limit is increased to 
$10^{-6}\msun$ if subject to cosmic ray heating similar to that in the 
Galactic disc for an interval $\ga kT/\Gamma\sim10^5 \u yr $.  As halo 
clouds take much longer than this to pass through the cosmic-ray disc, 
this limit is appropriate even for a halo cloud population.  For cloud 
masses in the range $10^{-6}$--$10^{-1.7}\;\msun$ the radiative cooling 
simply readjusts, on the timescale $\tkh$, to maintain equilibrium as 
$\Gamma$ varies through the orbit ($\sim10^8$~yr) of a cloud around 
the Galaxy.

The typical particle radius $a$ is constrained by the requirement that 
the clouds not produce significant extinction at optical wavelengths: 
the geometrical optical depth of the particles, $\tau_g \approx 
\tau/C_m a \rho_s $ should be less than 1.  Adopting $T=5$\,K and $\tau 
= 0.01$ this translates to $a(\u mm ) \ga (10^{-4}\u cm )/\lambda_2$.
Millimeter-size particles settle to the centre of the cloud in $\sim 
10^4 \u yr $, this time is shortened if $\lambda_2$ is 
significantly less than $10^{-4}\u cm $ and the particles are required to be 
larger.  Settling may be counteracted by convective motions or 
sublimation resulting from the higher temperatures deeper within the 
cloud --- issues that must await a more sophisticated treatment of the 
cloud structure.

If the hypothesised population of clouds exists, their thermal microwave
emission may be detectable as a Galactic continuum background at temperatures
just above the Cosmic Microwave Background; a Galactic component of this kind
has in fact been isolated in the COBE FIRAS data (Reach et al.\ 1985).  One would
like to compare these data with the theory presented here, but it is
difficult to predict the total microwave intensity for our model because
the distribution of cosmic-ray density away from the Galactic plane is
only loosely constrained (see Webber et al.\ 1994). A similar uncertainty afflicts
the modelling of $\gamma$-ray production from baryonic material in the
Galactic halo (cf. de Paolis et al.\ 1995; Salati et al.\ 1996; de Paolis 
et al.\ 1999; Kalberla, Shchekinov \& Dettmar 1999).
Nevertheless, microwave and $\gamma$-ray emissivities are each
proportional to the local cosmic-ray flux (assuming the cosmic-ray spectrum
does not vary greatly), so we can write $I_\mu\simeq I_\gamma j_\mu/
j_\gamma$, for emissivities $j$ and intensities $I$. At high latititudes
the Galactic $\gamma$-ray background is $I_\gamma\sim10^{-6}\;{\rm ph\,
cm^{-2}\,s^{-1}\,sr^{-1}}$, above 1~GeV (Dixon et al.\ 1998); local to the Sun
the corresponding (optically thin) emissivity is $1.1\times10^{-3}\;{\rm
ph\;s^{-1}\,g^{-1}\,sr^{-1}}$ (Bertsch et al.\ 1993). Thus for a cosmic-ray
heating rate (again, local to the Sun) of $\Gamma=4\pi j_\mu\sim3\times
10^{-4}\;{\rm erg\,s^{-1}\,g^{-1}}$ (Cravens \& Dalgarno 1978), we expect
$I_\mu\sim2\times10^{-8}\;{\rm erg\,cm^{-2}\,s^{-1}\,sr^{-1}}$.
This is roughly 1\% of that observed in the FIRAS cold component
at high latitude (Reach et al.\ 1985), so the microwave data do not 
exclude the possibility of a cold cloud population heated by cosmic 
rays.  Sciama (1999) proposed that all of the cold excess may be 
accounted for by cosmic-ray heating of cold clouds, but this appears 
to be based on an overestimate of the gamma-ray flux, and an 
underestimate of the high-latitude FIRAS flux.

\section{Conclusions}
We have demonstrated that, by virtue of the solid/gas phase transition
of hydrogen,  cold, planetary-mass Galactic gas clouds can be thermally
stable even when they are heated at a temperature-independent rate.
Our analysis applies to the present epoch, with the microwave background
temperature at $T_b < 3$~K; for background temperatures $T_b\ga 6$~K,
our model admits no stable mass range.  Consequently the longevity of
the clouds at redshifts $z\ga 1$ is problematic. We cannot, however,
hope to address this issue until a firm theoretical basis for the
formation of such clouds has been established.

\acknowledgements 

We thank Sterl Phinney for making available a copy of an unpublished 
preprint and Bruce Draine for thoughtful comments on the manuscript.  
The Special Research Centre for Theoretical Astrophysics is funded by 
the Australian Research Council under its Special Research Centres 
programme.

%

%


\begin{references}  

\reference{} Abgrall, H., Roueff, E. \& Viala, Y. 1982, A\&A Supp, 50, 505
\reference{} Bertsch, D. L., Dame, T. M., Fichtel, C. E., Hunter, S. D.,
             Sreekumar, P., Stacy, J. G. \& Thaddeus, P. 1993, ApJ, 416, 587
\reference{} Cravens, T. E. \& Dalgarno, A. 1978, ApJ, 219, 750
\reference{} De Paolis, F., Ingrosso, G., Jetzer, Ph. \& Roncadelli, M. 1995,
             A\&A, 295, 567
\reference{} De Paolis, F., Ingrosso, G., Jetzer, Ph. \& Roncadelli, M. 
             1999, ApJ, 510, L103
\reference{} Draine~B.T. 1998, ApJ, 509, L41
\reference{} Draine~B.T. \& Lee~H.M. 1985 ApJ 285, 89
\reference{} Dixon, D. D., Hartmann, D. H., Kolaczyk, E. D., Samimi, J.,
             Diehl, R., Kanbach, G., Mayer-Hasselwander, H. \& Strong, A. W. 
             1998, New Astronomy,3, 539
\reference{} Fiedler, R. L., Dennison, B., Johnston, K. J. \& Hewish, A. 1987,
             Nature, 326, 675
\reference{} Fiedler, R., Dennison, B., Johnston, K. J., Waltman, E. B. \&
             Simon, R. S. 1994, ApJ, 430, 581
\reference{} Gerhard, O. \& Silk, J. 1996, ApJ, 472, 34
\reference{} Gianturco, F. A., Giorgi, P. G., Berriche, H. \& Gadea, F. X.
             1996, A\&A Supp, 117, 377
\reference{} Henriksen, R. N. \& Widrow, L. M. 1995, ApJ, 441, 70
\reference{} Jochemsen~R., Berlinsky~A.J., Verspaandonk~F. \& Silvera~I.F.
             1978,  J. Low Temp. Phys 32, 185
\reference{} Kalberla, P. M. W., Shchekinov, Yu. A. \& Dettmar, R. J. 1999, 
             A\&A, 350, L9
\reference{} Paczy\'{n}ski, B. 1996, ARAA, 34, 419
\reference{} Pfenniger, D. \& Combes, F. 1994, A\&A, 285, 94
\reference{} Pfenniger, D., Combes, F. \& Martinet, L. 1994, A\&A, 285, 79
\reference{} Phinney E. S. 1985, preprint
\reference{} Reach, W. T., et al. 1995, ApJ, 451, 188
\reference{} Salati, P., Chardonnet, P., Luo, X. C., Silk, J. \& Taillet,R.
             1996, A\&A, 313, 1
\reference{} Schramm~D.N. \& Turner~M.S. 1998 Rev. Mod. Phys 70, 303
\reference{} Sciama~D.W. 1999, MNRAS, submitted (astro-ph/9906159)
\reference{} Souers, P. C. 1986, Hydrogen properties for fusion energy
             (Berkeley: University of California Press)
\reference{} Turner, J., Kirby-Docken, K. \& Dalgarno, A. 1977 ApJS 35, 281
\reference{} Walker, M. 1999, MNRAS, 308, 551
\reference{} Walker, M. \& Wardle, M. 1998, ApJ, 498, L125
\reference{} Webber, W. R. 1998, ApJ, 506, 329
\reference{} Webber, W. R., Binns, W. R., Crary, D. \& Westphall, M. 1994,
             ApJ, 429, 764
\end{references}
\end{document}